\documentclass[aps,prd,preprint,showpacs,floatfix,nofootinbib]{revtex4-1}

\usepackage{dcolumn}

\usepackage{amsmath, amsfonts, amssymb, mathrsfs, times, float, color, bm}
\usepackage{extarrows}
\usepackage{graphicx}
\usepackage{graphicx,epstopdf}
\usepackage{dcolumn}
\usepackage{bm}

\usepackage{exscale}
\usepackage{relsize}

\newcommand{\nn}{\nonumber}

\begin{document}

\title[]
{Gravitational Deflection of Massive Particles by a Schwarzschild Black Hole in Radiation Gauge}
\author{Zonghai Li}
\affiliation{School of Physical Science and Technology, Southwest Jiaotong University, Chengdu, 610031, China}
\author{Xia Zhou}
\affiliation{Physics and Space Science College, China West Normal University, Nanchong, 637009, China}
\author{Weijun Li}
\author{Guansheng He}
\email{Email: hgs@usc.edu.cn}
\affiliation{School of Mathematics and Physics, University of South China, Hengyang, 421001, China}
\date{\today}

\begin{abstract}
The exact metric of a Schwarzschild black hole in the true radiation gauge was recently reported. In this work, we base on this gravity and calculate the gravitational deflection of relativistic massive particles up to the fourth post-Minkowskian order. It is found that the result is consistent with the previous formulations for both the case of dropping the fourth-order contribution and the case of light deflection. Our result might be helpful for future high-accuracy observations.

\begin{description}
\item[Keywords]
Gravitational Deflection; Relativistic Massive Particles; Radiation Gauge; Schwarzschild Black Hole
\end{description}


\end{abstract}

\maketitle

\section{Introduction}
Gravitational lensing of light acts as one of the most powerful tools in astrophysics because of its extensive applications. Not to be forgotten, the deflection of light by the Sun provided one of the first tests of general relativity~\cite{DED1920,Will2015}. Not limited to light, the gravitational lensing of massive particles has attracted more and more attentions in recent years~\cite{AR2002,WS2004,BSN2007,PNHZ2014,Tsupko2014,LYJ2016,Jusufi2018}. In 2002, Accioly and Ragusa~\cite{AR2002} derived the Schwarzschild deflection angle of relativistic massive particles up to the third post-Minkowskian (PM) order, the second-order contribution of which was, however, different from the one proposed in Ref.~\cite{BSN2007}. Recently, He and Lin~\cite{LinHe2016} considered the gravitational deflection of relativistic massive particles and light caused by a moving Kerr-Newman source numerically, and found that the second-order Schwarzschild contribution was in agreement with the former. This consistency was further confirmed by the analytical calculation via an iterative technique~\cite{LinHe2017}. There were also other investigations devoted to the deflection of massive particles in static and spherically symmetric spacetimes~\cite{CG2018,PJ2018,JBGA2018}, with their results matching with the Accioly and Ragusa's proposal.

In 2011, Chen and Zhu~\cite{CZ2011} proposed a true radiation gauge serving as coordinate conditions to solve the Einstein field equations and thus to investigate gravitational energy and radiation:
\begin{equation}
g^{ij}\Gamma^\lambda_{ij}=0~. \label{RG}
\end{equation}
Here and thereafter, Latin indices run from 1 to 3, and Greek indices run from 0 to 3. The exact metric of a Schwarzschild spacetime in this radiation gauge was later derived in Ref.~\cite{LinJiang2017}.

Since significant differences between the true radiation gauge and the harmonic gauge $g^{\mu\nu}\Gamma^\lambda_{\mu\nu}\!=\!0$~\cite{Weinberg1972} exist, which leads to the difference between the radiation-gauge solution~\cite{LinJiang2017} and the Schwarzschild metric in harmonic coordinates, it is interesting to study the classical tests of general relativity in the gravity reported in Ref.~\cite{LinJiang2017}. In addition, with the consideration of the great progress achieved in astronomical observations~\cite{Perryman2001,Laskin2006,SN2009,Malbet2012}, it is necessary to investigate the high-order contributions to the observable relativistic effects. In this article, we calculate the gravitational deflection of relativistic test particles including light up to the 4PM order in this gravity, based on the iterative technique proposed in Ref.~\cite{LinHe2017}. Our discussions are constrained in the weak-field, small-angle, and thin-lens approximation.

The structure of this article is as follows. In Section~\ref{Geodesics}, we give a brief review of the metric of a Schwarzschild black hole in radiation gauge and calculate the weak-field equations of motion of test particles. Section~\ref{DeflectionAngle} presents the derivation for the gravitational deflection of relativistic massive particles up to the 4PM order, followed by a summary in Section~\ref{Summary}.

Throughout the paper, metric signature $(-,~+,~+,~+)$ and natural units in which $G = c = 1$ are used.

\section{Weak-field equations of motion in radiation gauge} \label{Geodesics}

\subsection{The metric for a Schwarzschild black hole in radiation gauge}
Let $(\bm{e}_1,~\bm{e}_2,~\bm{e}_3)$ be the orthonormal basis of a three-dimensional Cartesian coordinate system. The exact metric of a Schwarzschild black hole in the form of the coordinates~$(t,~X,~Y,~Z)$~in radiation gauge reads~\cite{LinJiang2017}:
\begin{equation}
ds^2=-\frac{1-\frac{M}{2R}}{1+\frac{3M}{2R}}dt^2+\left(1+\frac{3M}{2R}\right)^2d\bm{X}^2-\frac{1+\frac{3M}{4R}-\frac{9M^2}{8R^2}}{1-\frac{M}{2R}}\frac{M}{R}\frac{\left(\bm{X}\cdot d\bm{X}\right)^2}{R^2}~,  \label{ds}
\end{equation}
where $M$ denotes the rest mass of the black hole,~$\bm{X}=(X,~Y,~Z)$,~$R=\sqrt{X^2+Y^2+Z^2}$, and $\bm{X}\cdot d\bm{X}=XdX+XdX+ZdZ$~. $-M/R$ represents the Newtonian gravitational potential. Notice that the exact metric of a Schwarzschild black hole in harmonic coordinates $(t,~x,~y,~z)$ via the harmonic gauge can be comparatively written as follows~\cite{Weinberg1972}:
\begin{equation}
ds^2=-\frac{1-\frac{M}{r}}{1+\frac{M}{r}}dt^2+\left(1+\frac{M}{r}\right)^2d\bm{x}^2+\frac{1+\frac{M}{r}}{1-\frac{M}{r}}\frac{M^2}{r^2}\frac{\left(\bm{x}\cdot d\bm{x}\right)^2}{r^2}~,  \label{ds-2}
\end{equation}
where $r=|\bm{x}|=\sqrt{x^2+y^2+z^2}$ and $\bm{x}\cdot d\bm{x}=xdx+ydy+zdz$. We can see that the metric of a Schwarzschild black hole in harmonic coordinates is different in form from the one in radiation-gauge coordinates, due to the difference between the harmonic and radiation gauges.

For calculating the gravitational deflection of relativistic massive particles up to the 4PM order, we only need the weak-field form of the radiation-gauge metric (i.e., Eq.~\eqref{ds}), which can be expanded in the post-Minkowskian approximation~\cite{PW2014} as follows:
{\small\begin{eqnarray}
&&{g_{00}}=-1+\frac{2 M}{R}-\frac{3 M^2}{R^2}+\frac{9 M^3}{2 R^3}-\frac{27 M^4}{4 R^4}+\mathcal{O}(M^4)~,   \label{g00} \\
&&{g_{0i}}=0~,   \label{g0i} \\
&&{g_{ij}}=\left(1+\frac{3 M}{R}+\frac{9 M^2}{4 R^2}\right)\delta _{ij}-\left(\frac{M}{R}+\frac{5 M^2}{4 R^2}-\frac{M^3}{2 R^3}-\frac{M^4}{4 R^4}\right)\frac{{ X_i}{X_j}}{R^2}+\mathcal{O}(M^4)~, \label{gij}
\end{eqnarray}}
where~$\delta _{ij}$~is the Kronecker symbol. The inverse metric up to the 3PM order is also needed:
{\small\begin{eqnarray}
&&{g^{00}}=-1-\frac{2 M}{R}-\frac{M^2}{R^2}-\frac{M^3}{2 R^3}+\mathcal{O}(M^3)~,   \label{g00-Inv} \\
&&{g^{0i}}=0 ~,   \label{g0i-Inv} \\
&&{g^{ij}}=\left(1-\frac{3 M}{R}+\frac{27 M^2}{4 R^2}-\frac{27 M^3}{2 R^3}\right)\delta_{ij}+\left(\frac{M}{R}-\frac{15 M^2}{4 R^2}+\frac{9 M^3}{R^3}\right)\frac{{ X_i}{X_j}}{R^2}+\mathcal{O}(M^3)~.  \label{gij-Inv}
\end{eqnarray}}

\subsection{Geodesic equations of test particles}
For simplicity, we consider the propagation of test particles which are confined to the equatorial plane ($Z=\partial/\partial Z=0$) of the gravitational source. Based on Eqs.~\eqref{g00} - \eqref{gij-Inv}, we can obtain the nonvanishing Christoffel symbols, which are given in Appendix~\ref{AppendixA}. Thus, we can get the explicit forms of the equations of motion of test particles up to the 4PM order as follows:
{\small\begin{eqnarray}
&&0={\ddot{t}}+\frac{\left(\frac{2M}{R}-\frac{2M^2}{R^2}+\frac{7M^3}{2R^3}\right)\left(X\dot{X}+Y\dot{Y}\right)\dot{t}}{R^2}-\frac{5\,X\,\dot{t}\,\dot{X}M^4}{R^6}+\mathcal{O}(M^4)~,  \label{MEt} \\
&&\nn0={\ddot{X}}+\frac{\left[2R^2 \dot{t}^2-\left(2X^2\!+\!5Y^2\right)\dot{X}^2\right]\!X M}{2 R^5}-\frac{3Y^3\dot{X}\dot{Y}M}{R^5}-\frac{\left[10R^2\dot{t}^2\!-\!\left(2X^2+9Y^2\right)\dot{X}^2\right]X M^2}{2 R^6}  \\
&&\nn\hspace*{15pt}+\,\frac{\left(X^2+4Y^2\right)X\dot{Y}^2 M}{2R^5}-\frac{\left(5X^2-9Y^2\right)Y \dot{X}\dot{Y} M^2}{2R^6}+\frac{\left[63R^2\,\dot{t}^2-\left(7X^2+ 27\,Y^2\right)\dot{X}^2\right]X M^3}{4 R^7}  \\
&&\hspace*{15pt}-\frac{7XY^2\dot{Y}^2M^2}{2R^6}\!+\!\frac{\left(13X^2\!-\!27Y^2\right)\!Y\!\dot{X}\dot{Y}\!M^3}{4 R^7}\!-\!\frac{\left[324R^2\,\dot{t}^2\!-\!\left(20 X^2\!+\!81Y^2\right)\!\dot{X}^2\right]\!XM^4}{8 R^8}
\!+\!\mathcal{O}(M^4)~,~~~~~~  \label{MEx}  \\
&&\nn0={\ddot{Y}}+\frac{\left[2R^2\dot{t}^2+\left(4X^2+Y^2\right)\!\dot{X}^2\right]\!YM}{2 R^5}-\frac{3X^3\dot{X}\dot{Y}M}{R^5}-\frac{\left(10R^2\,\dot{t}^2+7X^2\dot{X}^2\right)YM^2}{2R^6}  \\
&&\nn\hspace*{15pt}-\frac{\left(5X^2+2Y^2\right)Y\dot{Y}^2M}{2R^5}+\frac{\left(9X^2-5Y^2\right)X\dot{X}\dot{Y}M^2}{2R^6}+\frac{\left(63R^2\dot{t}^2+20 X^2\dot{X}^2\right)YM^3}{4R^7}  \\
&&\hspace*{15pt}+\frac{\left(9X^2\!+\!2Y^2\right)Y\dot{Y}^2M^2}{2R^6}\!-\!\frac{\left(27X^2\!-\!13Y^2\right)\!X\dot{X}\dot{Y}M^3}{4R^7}\!-\!\frac{\left(324R^2\dot{t}^2\!+\!61X^2\dot{X}^2\right)\!YM^4}{8R^8}
+\mathcal{O}(M^4)~, ~~~~~~\label{MEy}
\end{eqnarray}}
where a dot denotes the derivative with respect to the parameter $p$ which describes the trajectory~\cite{WS2004}, and $\dot{Y}$ has been assumed to be of the order $\sim\frac{M}{R}$, as done in Ref.~\cite{LinHe2016}. Note that Eqs.~\eqref{MEt} - \eqref{MEy} denote respectively the $t$, $X$, and $Y$-component of geodesic equations and that the motion is restricted to the equatorial plane. In addition, it can be seen that the analytical forms of the equations of motion of test particles in radiation gauge (i.e., Eqs.~\eqref{MEt} - \eqref{MEy}) are different from that in harmonic coordinates~\cite{LinHe2016} for the Schwarzschild case. Actually, one can also obtain Eqs.~\eqref{MEt} - \eqref{MEy} via the Euler-Lagrange method, as shown in Appendix~\ref{AppendixB}.

\section{Fourth-order Schwarzschild deflection of relativistic massive particles in radiation gauge} \label{DeflectionAngle}
\begin{figure*}[t]
\begin{center}
  \includegraphics[width=14.5cm]{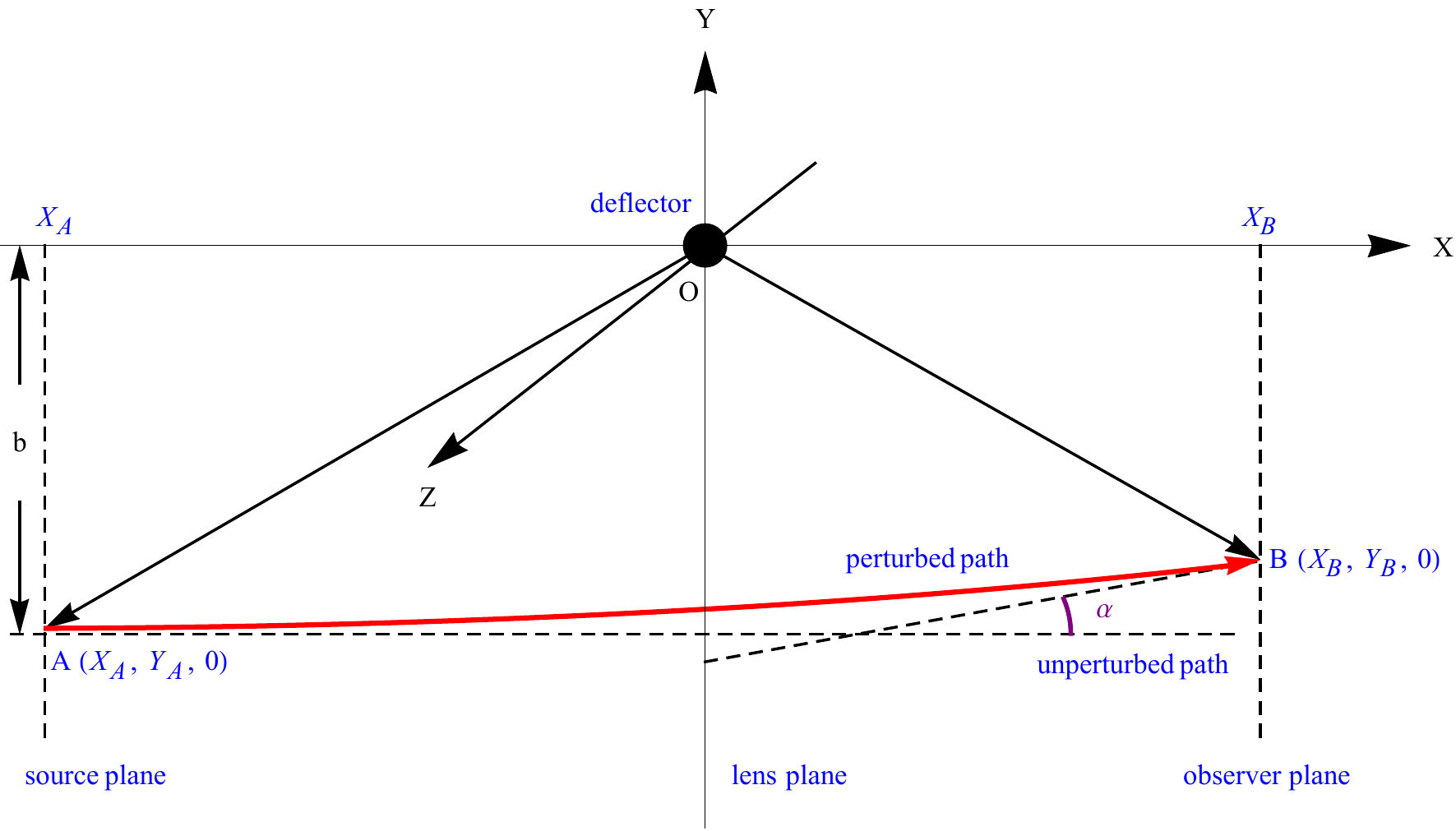}
  \caption{Schematic diagram for the gravitational deflection of a test particle due to a Schwarzschild source in radiation gauge. The gravitational deflection is greatly exaggerated to distinguish between the perturbed and unperturbed (dashed horizontal line) paths. }    \label{Figure1}
\end{center}
\end{figure*}
Let us consider the gravitational deflection of a massive particle caused by a Schwarzschild source in radiation gauge. The schematic diagram for the propagation of a test particle is given in Fig.~\ref{Figure1}. As done in Ref.~\cite{LinHe2017}, we assume the spatial coordinates of the source (denoted by $A$) and the detector (denoted by $B$) to be $(X_A,~Y_A,~0)$ and $(X_B,~Y_B,~0)$ respectively, with $X_A\ll-b$, $X_B\gg b$, and $Y_A\approx -b$. Here, $b$ is the impact parameter. The red line denotes the path of a test particle propagating from $X\rightarrow-\infty$ with a relativistic initial velocity
$\bm{w}|_{X\rightarrow-\infty}~(\approx \bm{w}|_{X\rightarrow X_A})~=w\bm{e}_1~$ ($0<w\leq1$).

The gravitational deflection angle of a test particle propagating from the source to the detector is defined as
{\small\begin{eqnarray}
\alpha\equiv  \arctan\left. \frac{dY}{dX}\right|_{B}-\arctan \left.\frac{dY}{dX}\right|_{A}=\arctan\left.\frac{\dot{Y}}{\dot{X}}\right|_{B}-\arctan\left.\frac{\dot{Y}}{\dot{X}}\right|_{A}~.\label{alpha}
\end{eqnarray}}
We assume the trajectory parameter $p$ to have the dimension of length~\cite{WS2004}, and adopt the iterative technique proposed in Ref.~\cite{LinHe2017} to derive the gravitational deflection up to the 4PM order.

First, Eqs.~\eqref{MEt} - \eqref{MEy} up to the 0PM order yield
\begin{eqnarray}
&&\dot{t}=\frac{1}{w}+\mathcal{O}(M^0)~, \label{0PMt} \\
&&\dot{X}=1+\mathcal{O}(M^0)~,           \label{0PMx} \\
&&\dot{Y}=0+\mathcal{O}(M^0)~,           \label{0PMy}
\end{eqnarray}
where the boundary conditions $\dot{t}|_{p\rightarrow -\infty}=\dot{t}|_{X\rightarrow -\infty}=\frac{1}{w}$, $\dot{X}|_{p\rightarrow -\infty}=\dot{X}|_{X\rightarrow -\infty}=1$,
and $\dot{Y}|_{p\rightarrow -\infty}=\dot{Y}|_{X\rightarrow -\infty}=0$ have been employed. Notice that Eqs.~\eqref{0PMt} - \eqref{0PMy} represent respectively the unperturbed forms of $\dot{t}$, $\dot{X}$, and $\dot{Y}$ without the existence of gravitational fields. Eqs.~\eqref{0PMx} - \eqref{0PMy} result in a 0PM parameter transformation and the 0PM form of $Y$:
\begin{eqnarray}
&&dX=\left[1+\mathcal{O}(M^0)\right]dp~, \label{ZO-PT} \\
&&Y=-b+\mathcal{O}(M^0)~,~~ \label{0th-y}
\end{eqnarray}
where the boundary condition $Y|_{p\rightarrow -\infty}=Y|_{X\rightarrow -\infty}=-b$ has been adopted. Note that Eq.~\eqref{0th-y} denotes the unperturbed form of the $Y$ coordinate of the test particle when there is no gravity.
Based on Eq.~\eqref{ZO-PT}, we then substitute Eqs.~\eqref{0PMt} - \eqref{0PMy} into Eqs.~\eqref{MEt} - \eqref{MEy} and integrate the latter equations over $p$ to obtain
\begin{eqnarray}
&&\dot{t}=\frac{1}{w}\left(1+\frac{2M}{\sqrt{X^2+b^2}}\right)+\mathcal{O}(M)~, \label{1th-dot-t} \\
&&\dot{X}=1-\left[1-\frac{1}{w^2}+\frac{b^2}{2\left(X^2+b^2\right)}\right]\frac{M}{\sqrt{X^2+b^2}}+\mathcal{O}(M)~, \label{1th-dot-x} \\
&&\dot{Y}=\left[\left(1+\frac{1}{w^2}\right)\left(1+\frac{X}{\sqrt{X^2+b^2}}\right)-\frac{b^2X}{2\left(X^2+b^2\right)^\frac{3}{2}}\right]\frac{M}{b}+\mathcal{O}(M)~.  \label{1th-dot-y}
\end{eqnarray}
With the help of Eqs.~\eqref{ZO-PT} and \eqref{0th-y}, we integrate Eq.~\eqref{1th-dot-y} over $p$ and have
\begin{eqnarray}
&&Y=-b+\left[\left(1+\frac{1}{w^2}\right)\left(\sqrt{X^2+b^2}+X\right)+\frac{b^2}{2\sqrt{X^2+b^2}}\right]\frac{M}{b}+\mathcal{O}(M)~.    \label{FO-y}
\end{eqnarray}
Eqs.~\eqref{1th-dot-t} - \eqref{FO-y} indicate that the first-order perturbations of the gravitational field on the analytical forms of $\dot{t}$, $\dot{X}$, $\dot{Y}$, and $Y$ have been considered.
In addition, Eq.~\eqref{1th-dot-x} yields the 1PM form of the parameter transformation
\begin{eqnarray}
&&dp=\left\{1+\left[1-\frac{1}{w^2}+\frac{b^2}{2\left(X^2+b^2\right)}\right]\frac{M}{\sqrt{X^2+b^2}}+\mathcal{O}(M)\right\}dX~,   \label{1PM-pt}
\end{eqnarray}
where the first-order perturbation of the gravitational field has been included.

Repeating the similar procedures, we take the second and third-order perturbations of the gravitational field into account, and thus obtain the explicit forms up to the 3PM order for $\dot{t},~\dot{X},~\dot{Y},~Y$, and the parameter transformation as follows:
\begin{eqnarray}
&&\nn\hspace*{5pt}\dot{t}=\frac{1}{w}\left(1+\frac{2M}{\sqrt{X^2+b^2}}\right)+\left[1+2\left(1+\frac{1}{w^2}\right)\left(1+\frac{X}{\sqrt{\,X^2+b^2}}\right)+\frac{b^2}{X^2+b^2}\right]\frac{M^2}{w\,\left({X^2+b^2}\right)} \\
&&\nn\hspace*{17pt}-\Bigg{[}\frac{2\left(1+\frac{1}{w^2}\right)^2\left(\sqrt{X^2+b^2}+X\right)}{X^2+b^2}
-\frac{\left(8+\frac{17}{w^2}+\frac{5}{w^4}\right)b^2+6\left(\frac{1}{4}+\frac{1}{w^2}\right)\left(\,\frac{\pi}{2}+\arctan\frac{X}{b}\,\right)b\,X}{\left(X^2+b^2\right)^{\frac{3}{2}}}  \\
&&\hspace*{17pt}-\frac{\left(5\!+\!\frac{9}{w^2}\!+\!\frac{4}{w^4}\right)\!b^2X}{\left(X^2+b^2\right)^2}+\frac{\left(\frac{1}{4}\!+\!\frac{3}{w^2}\!+\!\frac{3}{w^4}\right)b^4}{\left(X^2+b^2\right)^{\frac{5}{2}}}
\!-\!\frac{4\left(1\!+\!\frac{1}{w^2}\right)\!b^4\!X}{\left(X^2+b^2\right)^3}\!-\!\frac{5b^6}{4\left(X^2\!+\!b^2\right)^{\frac{7}{2}}}\Bigg{]}\!\frac{M^3}{w\,b^2}+\mathcal{O}(M^3)~, ~~~~~~~ \label{3th-dot-t}
\end{eqnarray}
\begin{eqnarray}
&&\nn\dot{X}=1-\frac{\left[\,1-\frac{1}{w^2}+\frac{b^2}{2\,\left(\,X^2\,+\,b^2\,\right)}\,\right]M}{\sqrt{X^2+b^2}}
-\Bigg{\{}\,\frac{\left(1+\frac{1}{w^2}\right)X\left[\,3\left(1+\frac{b^2}{X^2\,+\,b^2}\right)b^2+2\,\left(1+\frac{1}{w^2}\right)X^2\,\right]}{2\left(X^2+b^2\right)^{\frac{3}{2}}}  \\
&&\nn\hspace*{17pt}+\,\frac{b^2\,X^2\!\left[\,-\frac{5\,b^2}{4}+\frac{X^2}{w^4}+\left(2+\frac{5}{w^2}+\frac{1}{w^4}\right)\left(X^2+b^2\right)\,\right]}{\left(X^2+b^2\right)^3}
+\frac{\left(\frac{1}{2}+\frac{2}{w^2}\right)b^6+\left(1+\frac{1}{w^2}\right)^2X^6}{\left(X^2+b^2\right)^3}\Bigg{\}}\frac{M^2}{b^2}  \\
&&\nn\hspace*{17pt}-\,\Bigg{\{}\frac{3\left(1+\frac{5}{w^2}+\frac{4}{w^4}\right)\left(\frac{\pi}{2}+\arctan\frac{X}{b}\right)}{4}-\frac{\left(1+\frac{1}{w^2}\right)\left[\,2\left(5-\frac{12}{w^2}-\frac{8}{w^4}\right)b
-3\pi\left(1+\frac{4}{w^2}\right)\!X\,\right]}{8\sqrt{X^2+b^2}}  \\
&&\nn\hspace*{17pt}-\,\frac{\left(5\!-\!\frac{7}{w^2}\!-\!\frac{20}{w^4}\!-\!\frac{8}{w^6}\right)bX}{4\left(X^2+b^2\right)}
\!-\!\frac{b^2}{16\left(X^2\!+\!b^2\right)^\frac{3}{2}}\!\!\left[2\!\left(\!9\!+\!\frac{78}{w^2}\!+\!\frac{104}{w^4}\!+\!\frac{28}{w^6}\!\right)\!b-3\pi\!\left(\!1\!+\!\frac{2}{w^2}\!-\!\frac{8}{w^4}\!\right)\!X\right]  \\
&&\nn\hspace*{17pt}-\,\frac{\left(3\!+\!\frac{7}{w^2}\!+\!\frac{6}{w^4}\!+\!\frac{2}{w^6}\right)b^3X}{\left(X^2+b^2\right)^2}
\!+\!\frac{3\left[\,2\left(9\!+\!\frac{25}{w^2}\!+\!\frac{22}{w^4}\!+\!\frac{4}{w^6}\right)b\!+\!3\pi\left(1+\frac{4}{w^2}\right)\!X\,\right]\!b^4}{16\left(X^2+b^2\right)^\frac{5}{2}}
+\frac{3\left(1+\frac{4}{w^2}\right)\!X}{8\left(X^2+b^2\right)^\frac{5}{2}}  \\
&&\nn\hspace*{17pt}\times\!\left[6b^4\!+\!\left(\!5\!+\!\frac{2}{w^2}\!\right)\!b^2X^2\!+\!2\!\left(\!1\!+\!\frac{1}{w^2}\!\right)\!X^4\!\right]\!\arctan\frac{X}{b}
\!-\!\frac{\left(5\!-\!\frac{3}{w^2}\!-\!\frac{8}{w^4}\right)\!b^5\!X}{2\left(X^2+b^2\right)^3}\!-\!\frac{\left(77\!+\!\frac{70}{w^2}\!+\!\frac{60}{w^4}\right)\!b^7}{16\left(X^2+b^2\right)^{\frac{7}{2}}}  \\
&&\hspace*{17pt}
+\frac{9\left(1+\frac{1}{w^2}\right)b^7X}{2\left(X^2+b^2\right)^4}+\frac{21b^9}{16\left(X^2+b^2\right)^{\frac{9}{2}}} \Bigg{\}}\frac{M^3}{b^3}+\mathcal{O}(M^3)~,~~~~~\label{3th-dot-x}
\end{eqnarray}
\begin{eqnarray}
&&\nn\dot{Y}\!=\!\left[\left(1+\frac{1}{w^2}\right)\left(1+\frac{X}{\sqrt{X^2+b^2}}\right)-\frac{b^2X}{2\left(X^2+b^2\right)^\frac{3}{2}}\right]\!\frac{M}{b}
\!+\!\Bigg{\{}\frac{3}{4}\left(1+\frac{4}{w^2}\right)\!\left(\frac{\pi}{2}+\arctan\frac{X}{b}\right)  \\
&&\nn\hspace*{15pt}\!-\frac{\left(1\!-\!\frac{12}{w^2}\!-\!\frac{4}{w^4}\right)\!bX}{4\left(X^2+b^2\right)}
\!-\!\frac{\left(1\!+\!\frac{1}{w^2}\right)\!b\!\left[2b^2\!+\!\left(\!1\!-\!\frac{1}{w^2}\!+\!\frac{3b^2}{2\left(X^2+b^2\right)}\!\right)\!\!X^2\right]}{\left(X^2+b^2\right)^\frac{3}{2}}
\!+\!\frac{\left[1\!-\!\frac{1}{w^2}\!-\!\frac{3b^2}{4\left(X^2+b^2\right)}\right]\!b^3\!X}{\left(X^2+b^2\right)^2}\!\Bigg{\}}\!\frac{M^2}{b^2}  \\
&&\nn\hspace*{15pt}+\,\Bigg{\{}1\,+\,\frac{13}{w^2}\,+\,\frac{3}{w^4}-\frac{1}{w^6}\,+\,\frac{\left(\,2+\frac{6}{w^2}+\frac{7}{w^4}+\frac{3}{w^6}\,\right)b^2}{X^2+b^2}
\,-\,\frac{3\,\left(1+\frac{4}{w^2}\right)\,b\,\left(\,\frac{\pi}{2}+\arctan\frac{X}{b}\,\right)}{8\left(X^2+b^2\right)^\frac{3}{2}}  \\
&&\nn\hspace*{15pt}\!\times\!\!\left[4b^2\!+\!2\!\left(\!1\!-\!\frac{1}{w^2}\!\right)\!\!X^2\!+\!\frac{3b^2X^2}{X^2\!+\!b^2}\!\right]
\!+\!\frac{\left(13\!-\!\frac{31}{w^2}\!-\!\frac{52}{w^4}\!-\!\frac{8}{w^6}\right)\!b^4}{4(X^2+b^2)^2}\!-\!\frac{\left(37\!+\!\frac{21}{w^2}\!-\!\frac{16}{w^4}\right)\!b^6}{4\left(X^2+b^2\right)^3}
\!+\!\frac{9\!\left(1\!+\!\frac{1}{w^2}\right)\!b^8}{2\left(X^2\!+\!b^2\right)^4}
\end{eqnarray}
\begin{eqnarray}
&&\nn\hspace*{15pt}+\frac{X}{16\left(X^2\!+\!b^2\right)^{\frac{9}{2}}}\!\Bigg{[}2\!\left(\!7\!+\!\frac{138}{w^2}\!+\!\frac{68}{w^4}\!\right)\!b^8
\!+\!\left(\!41\!+\!\frac{1062}{w^2}\!+\!\frac{492}{w^4}\!+\!\frac{8}{w^6}\!\right)\!b^6\!X^2\!+\!2\!\left(41\!+\!\frac{787}{w^2}\!+\!\frac{342}{w^4}\right)     \\
&&\hspace*{15pt}\times\,b^4X^4\!+\!4\!\left(23\!+\!\frac{249}{w^2}\!+\!\frac{94}{w^4}\!-\!\frac{6}{w^6}\right)\!b^2X^6
\!+\!16\!\left(1\!+\!\frac{13}{w^2}\!+\!\frac{3}{w^4}\!-\!\frac{1}{w^6}\right)\!X^8\Bigg{]}\!\Bigg{\}}\frac{M^3}{b^3}\!+\!\mathcal{O}(M^3)~, \label{QM11}
\end{eqnarray}
\begin{eqnarray}
&&\nn Y=-\,b+\!\left[\left(\!1+\frac{1}{w^2}\!\right)\!\left(\sqrt{X^2+b^2}+X\right)+\frac{b^2}{2\sqrt{X^2\!+\!b^2}}\right]\!\frac{M}{b}
\!-\!\Bigg{\{}\!\left(\frac{1}{2}\!-\!\frac{1}{w^2}\right)^2\!b-3\left(\frac{1}{4}\!+\!\frac{1}{w^2}\right)\!X   \\
&&\nn \hspace*{15pt}\times\!\left(\frac{\pi}{2}\!+\!\arctan\frac{X}{b}\right)\!+\!\left(1\!+\!\frac{1}{w^2}\right)\!\!\left[1\!+\!\frac{1}{w^2}\!-\!\frac{b^2}{2\left(X^2\!+\!b^2\right)}\right]\!\frac{bX}{\sqrt{X^2+b^2}}
+\frac{\left[1\!-\!\frac{b^2}{2\left(X^2+b^2\right)}\right]\!b^3}{2(X^2+b^2)}\Bigg{\}}\frac{M^2}{b^2}\\
&&\nn\hspace*{15pt}+\Bigg{\{}\!\!-\frac{3\left(1+\frac{1}{w^2}\right)\left(1+\frac{4}{w^2}\right)b\left(\frac{\pi}{2}+\arctan\frac{X}{b}\right)}{4}
+\left(3+\frac{19}{w^2}+\frac{9}{w^4}+\frac{1}{w^6}\right)\left(\sqrt{X^2+b^2}+X\right)  \\
&&\nn\hspace*{15pt}-\frac{b}{16\sqrt{X^2+b^2}}\left[\,4\left(9+\frac{53}{w^2}+\frac{30}{w^4}+\frac{2}{w^6}\right)b+3\,\pi\left(3+\frac{14}{w^2}+\frac{8}{w^4}\right)X\,\right]
-\frac{3\,\left(1+\frac{4}{w^2}\right)b\,X}{8\sqrt{X^2+b^2}}  \\
&&\nn\hspace*{15pt}\times\!\left[\left(3+\frac{2}{w^2}\right)\arctan\frac{X}{b}-\left(1+\frac{b^2}{X^2+b^2}\right)\left(\frac{\pi}{2}+\arctan\frac{X}{b}\right)\right]
-\frac{3\left(1+\frac{5}{w^2}+\frac{4}{w^4}\right)b^2X}{4\left(X^2+b^2\right)}\\
&&\nn\hspace*{15pt}+\frac{\left(9\!+\!\frac{19}{w^2}\!+\!\frac{18}{w^4}\!+\!\frac{4}{w^6}\right)\!b^4}{8\left(X^2+b^2\right)^{\frac{3}{2}}}
\!-\!\frac{\left(1\!+\!\frac{1}{w^2}\right)\!\left(5\!-\!\frac{4b^2}{X^2+b^2}\right)\!b^4X}{4\left(X^2+b^2\right)^2}
\!-\!\frac{\left(19\!+\!\frac{18}{w^2}\!+\!\frac{12}{w^4}\right)\!b^6}{16\left(X^2+b^2\right)^\frac{5}{2}}\!+\!\frac{5b^8}{16\left(X^2\!+\!b^2\right)^\frac{7}{2}}\!\Bigg{\}}\frac{M^3}{b^3}  \\
&&\hspace*{15pt}+\,\mathcal{O}(M^3)~, ~~~~~~ \label{QM12}
\end{eqnarray}
\begin{eqnarray}
&&\nn{dp}=\!\Bigg{\{}\!1\!+\!\!\left[1\!-\!\frac{1}{w^2}+\frac{b^2}{2\left(X^2\!+\!b^2\right)}\right]\!\!\frac{M}{\sqrt{X^2\!+\!b^2}}
+\!\left\{\!\frac{\left(1\!+\!\frac{1}{w^2}\right)\!X\!\left[6b^4\!+\!\left(5\!+\!\frac{2}{w^2}\right)\!b^2X^2\!+\!2\left(1\!+\!\frac{1}{w^2}\right)\!X^4\right]}{2\left(X^2+b^2\right)^\frac{5}{2}} \right. \\
&&\nn\hspace*{14pt}\left. +\,\frac{\left(11-\frac{4}{w^2}+\frac{4}{w^4}\right)b^6+3\left(5+\frac{4}{w^4}\right)b^4X^2
+12\left(1+\frac{1}{w^2}+\frac{1}{w^4}\right)b^2\,X^4+4\left(1+\frac{1}{w^2}\right)^2X^6}{4\left(X^2+b^2\right)^3}\!\right\}\!\frac{M^2}{b^2}  \\
&&\nn\hspace*{15pt}+\Bigg{\{}\!\frac{3\pi\!\left(\!1\!+\!\frac{5}{w^2}\!+\!\frac{4}{w^4}\!\right)\!b^2\!\left(b^6\!\!+\!4b^4\!X^2\!\!+\!6b^2\!X^4\!\!+\!4X^6\right)}{8\left(X^2+b^2\right)^4}
\!+\!\frac{3\!\left(\!1\!+\!\frac{5}{w^2}\!+\!\frac{4}{w^4}\!\right)\!\left(2b\!+\!\pi X\right)\!\left(\sqrt{\!X^2\!+\!b^2}\!+\!X\right)\!X^7}{8\left(X^2+b^2\right)^\frac{9}{2}}  \\
&&\nn\hspace*{15pt}+\frac{3\pi\,b^2X\!\left[6\left(1\!+\!\frac{4}{w^2}\right)\!b^6\!+\!\left(17+\frac{70}{w^2}+\frac{8}{w^4}\right)\!b^4\!X^2\!+\!6\left(3+\frac{13}{w^2}+\frac{4}{w^4}\right)\!b^2\!X^4
\!+\!3\!\left(3+\frac{14}{w^2}+\frac{8}{w^4}\right)\!X^6\right]}{16\left(X^2+b^2\right)^\frac{9}{2}} \\
&&\nn\hspace*{15pt}+\frac{\left(19-\frac{38}{w^2}-\frac{24}{w^4}-\frac{8}{w^6}\right)\!b^9}{8\left(X^2+b^2\right)^{\frac{9}{2}}}+\frac{\left(1+\frac{1}{w^2}\right)\!b^3X\left[3\left(9-\frac{4}{w^2}\right)b^4
+\left(5-\frac{12}{w^2}\right)\!b^2X^2+\left(5+\frac{12}{w^2}\right)\!X^4\right]}{4\left(X^2+b^2\right)^4}  \\
&&\nn\hspace*{15pt}+\frac{\left[\,3\left(19\!-\!\frac{38}{w^2}\!-\!\frac{28}{w^4}\!-\!\frac{24}{w^6}\right)b^4+6\left(12\!-\!\frac{5}{w^2}\!-\!\frac{6}{w^4}\!-\!\frac{16}{w^6}\right)b^2X^2
+2\left(15\!+\!\frac{34}{w^2}\!+\!\frac{24}{w^4}\!-\!\frac{20}{w^6}\right)X^4\,\right]b^3X^2}{16(X^2+b^2)^\frac{9}{2}}  \\
&&\hspace*{15pt}+\frac{3\left(1\!+\!\frac{4}{w^2}\right)\left[1+\frac{1}{w^2}+\frac{\left[6b^4+\left(5+\frac{2}{w^2}\right)b^2X^2+2\left(1+\frac{1}{w^2}\right)X^4\right]X}
{2\left(X^2+b^2\right)^{\frac{5}{2}}}\right]\!\arctan\frac{X}{b}}{4}\Bigg{\}}\frac{M^3}{b^3}+\mathcal{O}(M^3)\Bigg{\}}dX~.~~~~~~  \label{pt3PM}
\end{eqnarray}

We finally substitute Eqs.~\eqref{3th-dot-t} - \eqref{pt3PM} into the integration of Eq.~\eqref{MEy} over $p$ and get the explicit form of $\dot{Y}$ up to the 4PM order
{\small\begin{eqnarray}
\nn&&\dot{Y}=\left[\!\left(1\!+\!\frac{1}{w^2}\right)\!\!\left(1\!+\!\frac{X}{\sqrt{\!X^2\!+\!b^2}}\right)\!-\!\frac{b^2X}{2\left(X^2\!+\!b^2\right)^{\frac{3}{2}}}\!\right]\!\frac{M}{b}
\!+\!\Bigg{\{}\!\frac{3\left(1\!+\!\frac{4}{w^2}\right)\!\left(\frac{\pi}{2}\!+\!\arctan\frac{X}{b}\right)}{4}\!-\!\frac{\left(1\!-\!\frac{12}{w^2}\!-\!\frac{4}{w^4}\right)\!bX}{4\left(X^2+b^2\right)}  \\
&&\nn\hspace*{13pt}-\frac{\left(1+\frac{1}{w^2}\right)b}{\left(X^2+b^2\right)^\frac{3}{2}}\left\{2\,b^2+\left[1\!-\!\frac{1}{w^2}\!+\!\frac{3b^2}{2\left(X^2+b^2\right)}\right]\!X^2\right\}
+\frac{b^3X}{\left(X^2+b^2\right)^2}\!\left[1-\frac{1}{w^2}-\frac{3b^2}{4\left(X^2+b^2\right)}\right]\!\Bigg{\}}\frac{M^2}{b^2}  \\
&&\nn\hspace*{13pt}+\Bigg{\{}1\!+\!\frac{13}{w^2}\!+\!\frac{3}{w^4}\!-\!\frac{1}{w^6}\!+\!\frac{\left(2+\frac{6}{w^2}+\frac{7}{w^4}+\frac{3}{w^6}\right)b^2}{X^2+b^2}
+\frac{\left(13-\frac{31}{w^2}-\frac{52}{w^4}-\frac{8}{w^6}\right)b^4}{4(X^2+b^2)^2}-\frac{\left(37+\frac{21}{w^2}-\frac{16}{w^4}\right)b^6}{4\left(X^2+b^2\right)^3}  \\
&&\nn\hspace*{13pt}+\,\frac{9\left(\,1+\frac{1}{w^2}\,\right)\,b^8}{2\left(X^2+b^2\right)^4}
+\frac{X}{16\,\left(X^2+b^2\right)^\frac{9}{2}}\Bigg{[}\,2\,\left(7+\frac{138}{w^2}+\frac{68}{w^4}\right)b^8+\left(41+\frac{1062}{w^2}+\frac{492}{w^4}+\frac{8}{w^6}\right)\,b^6\,X^2  \\
&&\nn\hspace*{13pt}+2\!\left(41+\frac{787}{w^2}+\frac{342}{w^4}\right)\!b^4X^4+4\!\left(23+\frac{249}{w^2}+\frac{94}{w^4}-\frac{6}{w^6}\right)\!b^2X^6
+16\left(1+\frac{13}{w^2}+\frac{3}{w^4}-\frac{1}{w^6}\right)\!X^8\Bigg{]} \\
&&\nn\hspace*{13pt}-\frac{3\left(1+\frac{4}{w^2}\right)\!\left(\frac{\pi}{2}+\arctan\frac{X}{b}\right)b
\left[4b^2\!+\!2\left(1\!-\!\frac{1}{w^2}\right)\!X^2\!+\!\frac{3b^2X^2}{X^2+b^2}\right]}{8\left(X^2+b^2\right)^{\frac{3}{2}}}\Bigg{\}}\frac{M^3}{b^3}
+\Bigg{\{}\frac{3\left(19+\frac{464}{w^2}+\frac{416}{w^4}-\frac{64}{w^6}\right)}{64}  \\
&&\nn\hspace*{13pt}\times\!\left(\frac{\pi}{2}+\arctan\frac{X}{b}\right)+\frac{\left(121+\frac{880}{w^2}+\frac{1888}{w^4}-\frac{192}{w^6}-\frac{192}{w^8}\right)bX}{64\left(X^2+b^2\right)}
-\frac{\left(5-\frac{112}{w^2}-\frac{1680}{w^4}-\frac{1184}{w^6}-\frac{224}{w^8}\right)b^3X}{32\left(X^2+b^2\right)^2}  \\
&&\nn\hspace*{15pt}+\frac{\left(\,261+\frac{360}{w^2}-\frac{612}{w^4}-\frac{608}{w^6}-\frac{64}{w^8}\,\right)b^5X}{16\left(X^2+b^2\right)^3}
-\frac{3\left(\,97+\frac{142}{w^2}+\frac{50}{w^4}-\frac{32}{w^6}\,\right)b^7X}{8\left(X^2+b^2\right)^4}+\frac{\left(\,87+\frac{112}{w^2}+\frac{72}{w^4}\,\right)b^9X}{4\left(X^2+b^2\right)^5}  \\
&&\nn\hspace*{13pt}-\,\frac{1}{16\,\left(\,X^2+b^2\,\right)^{\frac{11}{2}}}\,
\left[\,32\,\left(\,1+\frac{17}{w^2}+\frac{8}{w^4}\,\right)b^{11}+2\,\left(\,57+\frac{1179}{w^2}+\frac{254}{w^4}-\frac{132}{w^6}\,\right)b^9\,X^2+b^7X^4 \right. \\
&&\nn\hspace*{13pt}\left.\times\left(161+\frac{3787}{w^2}-\frac{402}{w^4}-\frac{884}{w^6}+\frac{8}{w^8}\right)
+2\,\left(143+\frac{1564}{w^2}-\frac{713}{w^4}-\frac{534}{w^6}+\frac{32}{w^8}\right)b^5X^6+2\,b^3X^8  \right. \\
&&\nn\hspace*{13pt}\left.\times\left(7+\frac{583}{w^2}-\frac{412}{w^4}-\frac{200}{w^6}+\frac{52}{w^8}\right)
-4\,\left(1-\frac{50}{w^2}+\frac{13}{w^4}-\frac{12}{w^6}-\frac{12}{w^8}\right)b\,X^{10}\,\right]-\frac{5\,b^{11}X}{2\left(\,X^2+b^2\,\right)^{\,6}}  \\
&&\nn\hspace*{13pt}+\frac{3\left(1+\frac{4}{w^2}\right)\left(\frac{\pi}{2}+\arctan\frac{X}{b}\right)b^2\!
\left[\,2\left(2+\frac{4}{w^2}+\frac{3}{w^4}\right)+\frac{\left(13\,-\,\frac{44}{w^2}\,-\,\frac{8}{w^4}\right)b^2}{2\left(X^2+b^2\right)}
-\frac{\left(37\,-\,\frac{16}{w^2}\right)b^4}{2\,\left(X^2+b^2\right)^2}+\frac{9\,b^6}{\left(X^2\,+\,b^2\right)^3}\right]}{8(X^2+b^2)}  \\
&&\hspace*{13pt}-\frac{3\!\left(1\!+\!\frac{5}{w^2}\!+\!\frac{4}{w^4}\right)\!X\!\left(\frac{\pi}{2}\!+\!\arctan\frac{X}{b}\right)\!\!
\left[\!\left(7\!+\!\frac{2}{w^2}\right)\!b^6\!+\!25b^4\!X^2\!+\!5b^2\!X^4\!+\!2\left(1\!+\!\frac{1}{w^2}\right)\!X^6\right]}{8\left(X^2+b^2\right)^{\frac{7}{2}}}\!\!\Bigg{\}}\!\frac{M^4}{b^4}\!+\!\mathcal{O}(M^4)~.
~~~~~~~ \label{4th-dot-y}
\end{eqnarray}}

Hence, the 4PM gravitational deflection angle of a relativistic massive particle due to a Schwarzschild black hole in radiation gauge can be achieved by substituting Eqs.~\eqref{3th-dot-x}~and~\eqref{4th-dot-y}~into Eq.~\eqref{alpha} as follows:
\begin{eqnarray}
\alpha=\frac{2\left(1\!+\!\frac{1}{w^2}\right)\!M}{b}\!+\!\frac{3\left(1\!+\!\frac{4}{w^2}\right)\!\pi M^2}{4b^2}\!+\!\frac{2\left(5\!+\!\frac{45}{w^2}\!+\!\frac{15}{w^4}\!-\!\frac{1}{w^6}\right)\!M^3}{3b^3}
\!+\!\frac{105\left(\frac{1}{16}\!+\!\frac{1}{w^2}\!+\!\frac{1}{w^4}\right)\!\pi M^4}{4b^4}~,~~~~ \label{SB-angle1}
\end{eqnarray}
where the conditions $X_A\ll -b$ and $X_B\gg b$ have been used.

It is found that when the fourth-order contribution is dropped, Eq.~\eqref{SB-angle1} matches well with the result for the Schwarzschild deflection angle of a relativistic massive particle given in Ref.~\cite{AR2002}, which is
\begin{eqnarray}
\alpha=2\left(1+\frac{1}{w^2}\right)\frac{M}{b}+\frac{3 \pi}{4}\left(1+\frac{4}{w^2}\right)\frac{M^2}{b^2}+\frac{2}{3}\left(5+\frac{45}{w^2}+\frac{15}{w^4}-\frac{1}{w^6}\right)\frac{M^3}{b^3}~.~~~~ \label{SB-angle1-2}
\end{eqnarray}
Up to the 2PM order, Eq.~\eqref{SB-angle1} is also consistent with the results presented in Refs.~\cite{LinHe2017,CG2018,PJ2018,JBGA2018} in the framework of general relativity.
Moreover, for the case of $w=c=1$, Eq.~\eqref{SB-angle1} can be simplified to the 4PM deflection angle of light~\cite{KP2005}:
\begin{equation}
\alpha=\frac{4M}{b}+\frac{15\pi}{4}\frac{M^2}{b^2}+\frac{128}{3}\frac{M^3}{b^3}+\frac{3465\pi}{64}\frac{M^4}{b^4}~. \label{SB-angle2}
\end{equation}

Finally, it should be pointed out that the equations of motion and detailed processes for calculating the Schwarzschild deflection of test particles including light in radiation gauge are different from that in harmonic gauge~\cite{LinHe2017} in form. However, we can see that the gravitational deflection angle of a relativistic test particle caused by a Schwarzschild black hole is independent on the concrete gauges used in the derivation, such as the radiation gauge here or the harmonic gauge in Ref.~\cite{LinHe2017}. Since different gauge leads to different coordinates for a given geometry, we actually verify further that the gravitational deflection angle of a test particle is independent on concrete coordinates, including the radiation-gauge coordinates~\cite{CZ2011,LinJiang2017}.

\section{Summary}  \label{Summary}
In this work, we have applied an iterative technique to deriving the equatorial deflection of a relativistic massive particle up to the fourth post-Minkowskian order caused by a Schwarzschild black hole in radiation gauge. The fourth-order contribution to the Schwarzschild deflection angle of the massive particle is obtained for the first time. The resulting bending angle is consistent with that in Ref.~\cite{AR2002} when the fourth-order contribution is dropped, and that in Ref.~\cite{KP2005} for the case of light. Our result might be helpful for future high-accuracy observations.

\section*{Acknowledgements}
This work was supported in part by the National Natural Science Foundation of China (Grant No. 11647314), the Research Foundation of Education Department of Hunan Province (Grant No. 18C0427), and the Fundamental Research Funds of China West Normal University (Grant No. 18Q067).

\appendix

\section{Nonvanishing components of Christoffel symbols} \label{AppendixA}
The nonvanishing components of the Christoffel symbol can be derived directly as follows:
{\small\begin{eqnarray}
&&\Gamma_{01}^0={\Gamma_{10}^0}=\frac{X}{R^2}\left(\frac{M}{R}-\frac{M^2}{R^2}+\frac{7M^3}{4R^3}-\frac{5M^4}{2R^4}\right)~,  \\
&&\Gamma_{02}^0={\Gamma_{20}^0}=\frac{Y}{R^2}\left(\frac{M}{R}-\frac{M^2}{R^2}+\frac{7M^3}{4R^3}-\frac{5M^4}{2R^4}\right)~,  \\
&&\Gamma_{00}^1=\frac{X}{R^2}\left(\frac{M}{R}-\frac{5M^2}{R^2}+\frac{63M^3}{4R^3}-\frac{81M^4}{2R^4}\right)~,  \\
&&\Gamma_{11}^1=-\frac{X}{R^2}\!\left[\frac{\left(2X^2\!+\!5Y^2\right)M}{2R^3}\!-\!\frac{\left(2X^2\!+\!9Y^2\right)M^2}{2R^4}
\!+\!\frac{\left(7X^2\!+\!27Y^2\right)\!M^3}{4R^5}\!-\!\frac{\left(20X^2\!+\!81Y^2\right)\!M^4}{8R^6}\right]~,~~~~\\
&&\Gamma_{12}^1={\Gamma_{21}^1}=\frac{Y}{R^2}\left[-\frac{3Y^2M}{2R^3}\!-\!\frac{\left(5X^2\!-\!9Y^2\right)M^2}{4R^4}
\!+\!\frac{\left(13X^2\!-\!27Y^2\right)M^3}{8R^5}\!-\!\frac{\left(41X^2\!-\!81Y^2\right)M^4}{16R^6}\right]~,~~\\
&&\Gamma_{22}^1=\frac{X}{R^2}\left[\frac{\left(X^2+4Y^2\right)M}{2R^3}-\frac{7Y^2M^2}{2R^4}+\frac{5Y^2M^3}{R^5}-\frac{61Y^2 M^4}{8R^6}\right]~,\\
&&\Gamma_{00}^2=\frac{Y}{R^2}\left(\frac{M}{R}-\frac{5M^2}{R^2}+\frac{63M^3}{4R^3}-\frac{81 M^4}{2R^4}\right),  \\
&&\Gamma_{11}^2=\frac{Y}{R^2}\left[\frac{\left(4X^2+Y^2\right)M}{2R^3}-\frac{7X^2M^2}{2R^4}+\frac{5X^2M^3}{R^5}-\frac{61X^2M^4}{8R^6}\right],\\
&&\Gamma_{12}^2={\Gamma_{21}^2}=\frac{X}{R^2}\!\left[-\frac{3X^2M}{2R^3}\!+\!\frac{\left(9X^2\!-\!5Y^2\right)M^2}{4R^4}
\!-\!\frac{\left(27X^2\!-\!13Y^2\right)M^3}{8R^5}\!+\!\frac{\left(81X^2\!-\!41Y^2\right)M^4}{16R^6}\right]~,~~\\
&&\hspace*{-1pt}\Gamma_{22}^2=-\frac{Y}{R^2}\!\left[\frac{\left(5X^2\!+\!2Y^2\right)M}{2R^3}\!-\!\frac{\left(9X^2\!+\!2Y^2\right)M^2}{2R^4}\!+\!\frac{\left(27X^2\!+\!7Y^2\right)\!M^3}{4R^5}
\!-\!\frac{\left(81X^2\!+\!20Y^2\right)\!M^4}{8R^6}\right]~.~~~~~~~~~~
\end{eqnarray}}

\section{4PM geodesic equations based on the Euler-Lagrange method} \label{AppendixB}
The Lagrangian function is defined as~\cite{WS2004}:
{\small\begin{eqnarray}
&&L=-g_{\mu\nu}\frac{dX^{\mu}}{dp}\frac{dX^{\nu}}{dp}~,  \label{B-L}
\end{eqnarray}}
which satisfies the Euler-Lagrange equation
{\small\begin{eqnarray}
&&\frac{d}{dp}\frac{\partial L}{\partial \dot{X}^{\mu}}-\frac{\partial L}{\partial X^{\mu}}=0~, \label{E-L}
\end{eqnarray}}
where a dot denotes the derivative with respect to $p$\,.

For test particles propagating in the equatorial plane, considering Eqs.~\eqref{g00} - \eqref{gij} and substituting Eq.~\eqref{B-L}~into Eq.~\eqref{E-L}, we can derive the nonzero components of the geodesic equations up to the 4PM order as follows:
{\small\begin{eqnarray}
\hspace*{-0.6cm}&&0={\ddot{t}}+\left(\frac{2M}{R}-\frac{2M^2}{R^2}+\frac{7M^3}{2R^3}\right)\frac{\left(X\dot{X}+Y\dot{Y}\right)\dot{t}}{R^2}-\frac{5X\dot{t}\dot{X}M^4}{R^6}+O(M^4)~,  \label{B-t}  \\
\hspace*{-0.6cm}&&\nn0\,=\,{\ddot{X}}+\frac{\left[\,2\,R^2\,\dot{t}^{\,2}-\left(2\,X^2+5\,Y^2\right)\dot{X}^2\,\right]XM}{2R^5}-\frac{3\,Y^3\,\dot{X}\,\dot{Y}M}{R^5}
+\frac{\left(X^2+4\,Y^2\right)X\,\dot{Y}^2\,M}{2R^5}-\frac{XM^2}{2R^8} \\
\hspace*{-0.6cm}&&\nn\times\left[\,2\,R^2\left(5\,X^2+6\,Y^2\right)\dot{t}^2-\left(2\,X^4+7\,X^2\,Y^2+8\,Y^4\right)\dot{X}^2\,\right]+\frac{\left(X^4+4\,X^2\,Y^2+9\,Y^2\right)Y \dot{X}\,\dot{Y} M^2}{2R^8} \\
&&\nn+\,\frac{\left[\,2\,R^2\left(63\,X^4+149\,X^2\,Y^2+90\,Y^4\right)\dot{t}^2-\left(14\,X^6+42\,X^4\,Y^2+63\,X^2Y^4+47\,Y^6\right)\dot{X}^2\,\right]\!X M^3}{8R^{11}} \\
&&\nn-\,\frac{\left(2\,X^2+5\,Y^2\right)X\,Y^2\,\dot{Y}^2\,M^2}{2R^8}-\frac{\left(\,14\,X^6+30\,X^4\,Y^2+31\,X^2Y^4+27\,Y^6\,\right)\,Y\,\dot{X}\,\dot{Y}\,M^3}{4R^{11}}-\frac{XM^4}{8R^{14}} \\
&&\nn\times\!\!\left[4\!\left(81X^6\!\!+\!282X^4Y^2\!\!+\!334X^2Y^4\!\!+\!135Y^6\right)\!R^2\dot{t}^2\!\!-\!\left(20X^8\!\!+\!80X^6Y^2\!\!+\!129X^4Y^4\!\!+\!128X^2Y^6\!\!+\!71Y^8\right)\!\dot{X}^2\!\right] \\
&&-\frac{XYM}{R^3}\!\left[1-\frac{\left(3X^2+7Y^2\right)M}{4R^3}+\frac{Y^2\left(3X^2+5Y^2\right)M^2}{2R^6}\right]\ddot{Y}+O(M^4)~,  \label{B-X}  \\
&&\nn 0\,=\,\ddot{Y}+\frac{\left[\,2\,R^2\,\dot{t}^2+\left(\,4\,X^2+Y^2\,\right)\dot{X}^2\,\right]Y M}{2R^5}-\frac{3\,X^3\,\dot{X}\,\dot{Y}\,M}{R^5}
-\frac{\left(5\,X^2+2\,Y^2\right)Y\,\dot{Y}^{2}\,M}{2R^5}  \\
&&\nn-\frac{\left[2\left(6X^2\!+\!5Y^2\right)R^2\dot{t}^2\!+\!\left(5X^2\!+\!2Y^2\right)X^2\dot{X}^2\right]Y\!M^2}{2R^8}+\frac{\left(9X^4+4X^2Y^2+Y^4\right)X\dot{X}\dot{Y}M^2}{2R^8}  \\
&&\nn+\,\frac{\left[\,2\left(90\,X^4+149\,X^2\,Y^2+63\,Y^4\right)\,R^2\,\dot{t}^2+\left(18\,X^4-5\,X^2\,Y^2-11\,Y^4\right)\,X^2\,\dot{X}^2\,\right]YM^3}{8R^{11}}  \\
&&\nn+\frac{\left(8X^4+7X^2Y^2+2Y^4\right)Y\dot{Y}^2M^2}{2R^8}-\frac{\left(27X^6+31X^4Y^2+30X^2Y^4+14Y^6\right)X\dot{X}\dot{Y}M^3}{4R^{11}}  \\
&&\nn-\frac{Y\!\left[4\!\left(135X^6\!+\!334X^4Y^2\!+\!282X^2Y^4\!+\!81Y^6\right)\!R^2\,\dot{t}^2\!+\!\left(13X^6\!\!-\!44X^4Y^2\!\!-\!65X^2Y^4\!\!-\!20Y^6\right)\!X^2\dot{X}^2\right]\!M^4}{8R^{14}}  \\
&&-\frac{XYM}{R^3}\!\left[1-\frac{\left(7X^2+3Y^2\right)M}{4R^3}+\frac{\left(5X^2+3Y^2\right)X^2M^2}{2R^6}\right]\ddot{X}+O(M^4)~.  \label{B-Y}
\end{eqnarray}}
Now we adopt an iterative technique to calculate the explicit forms up to the 3PM order for $\ddot{Y}$ and $\ddot{X}$ in Eqs.~\eqref{B-X} - \eqref{B-Y}, respectively. First, Eqs.~\eqref{B-X}~and~\eqref{B-Y} up to the 1PM order yield respectively
{\small\begin{eqnarray}
&&\ddot{X}=-\frac{\left[2 R^2 \dot{t}^2-\left(2 X^2+ 5 Y^2\right)\dot{X}^2\right]X M}{2 R^5}+O(M)~, \label{B-X-1} \\
&&\ddot{Y}=-\frac{\left[2R^2\dot{t}^2+\left(4X^2+Y^2\right)\dot{X}^2\right]Y M}{2 R^5}+O(M)~.  \label{B-Y-1}
\end{eqnarray}}
We then substitute Eqs.~\eqref{B-Y-1} and~\eqref{B-X-1} into Eqs.~\eqref{B-X} and~\eqref{B-Y}, respectively, and get the explicit forms of $\ddot{X}$ and $\ddot{Y}$ up to the 2PM order
{\small\begin{eqnarray}
&&\ddot{X}\!=\!-\frac{\left[2R^2\dot{t}^2\!-\!\left(2X^2\!+\!5Y^2\right)\!\dot{X}^2\right]\!\!X\!M}{2R^5}\!+\!\frac{3Y^3\dot{X}\dot{Y}\!M}{R^5}
\!+\!\frac{\left[10R^2\dot{t}^2\!-\!\left(2X^2\!+\!9Y^2\right)\!\dot{X}^2\right]\!\!X\!M^2}{2R^6}\!+\!O(M^2)~,~~~~~~~~ \label{B-X-2}  \\
&&\ddot{Y}\!=\!-\frac{\left[2R^2\dot{t}^2+\left(4X^2+Y^2\right)\dot{X}^2\right]\!Y\!M}{2 R^5}\!+\!\frac{3X^3\dot{X}\dot{Y}\!M}{R^5}\!+\!\frac{\left(10R^2\dot{t}^2+7X^2\dot{X}^2\right)\!Y\!M^2}{2R^6}\!+\!O(M^2)~. \label{B-Y-2}~
\end{eqnarray}}
Similarly, by plugging Eqs.~\eqref{B-Y-2} and \eqref{B-X-2} into Eqs.~\eqref{B-X} and~\eqref{B-Y} respectively, we obtain
{\small\begin{eqnarray}
&&\nn \ddot{X}=-\frac{\left[2 R^2 \dot{t}^2-\left(2X^2+5Y^2\right)\dot{X}^2\right]\!XM}{2R^5}+\frac{3Y^3\dot{X}\dot{Y}M}{R^5}+\frac{\left[10R^2\dot{t}^2-\left(2X^2+9Y^2\right)\dot{X}^2\right]\!XM^2}{2R^6}  \\
&&\hspace*{8pt}-\frac{\left(X^2\!+\!4Y^2\right)\!X\dot{Y}^2\!M}{2R^5}\!+\!\frac{\left(5X^2\!-\!9Y^2\right)\!Y\!\dot{X}\dot{Y}\!M^2}{2R^6}
\!-\!\frac{\left[63R^2\dot{t}^2\!-\!\left(7X^2\!+\!27Y^2\right)\!\dot{X}^2\right]\!X\!M^3}{4R^7}\!+\!O(M^3)~,~~~~~~~~~~ \label{B-X-3}  \\
&&\nn \ddot{Y}=-\,\frac{\left[\,2\,R^2\,\dot{t}^2+\left(4X^2+Y^2\right)\dot{X}^2\,\right]Y M}{2 R^5}+\frac{3\,X^3\,\dot{X}\,\dot{Y}M}{R^5}+\frac{\left(10\,R^2\,\dot{t}^2+7\,X^2\,\dot{X}^2\right)YM^2}{2R^6}  \\
&&\hspace*{8pt}+\frac{\left(5X^2\!+\!2Y^2\right)Y\dot{Y}^2\!M}{2R^5}\!-\!\frac{\left(9X^2\!-\!5Y^2\right)X\dot{X}\dot{Y}\!M^2}{2R^6}
\!-\!\frac{\left(63R^2\,\dot{t}^2\!+\!20 X^2\dot{X}^2\right)Y\!M^3}{4R^7}\!+\!O(M^3)~.   \label{B-Y-3}
\end{eqnarray}}
Finally, the explicit forms of the $X$- and $Y$- components of the geodesic equations up to the 4PM order can be computed on the basis of Eqs.~\eqref{B-Y-3} and~\eqref{B-X-3} respectively as follows:
{\small\begin{eqnarray}
&&\nn0={\ddot{X}}\!+\!\frac{\left[2R^2\dot{t}^2\!-\!\left(2X^2\!+\!5Y^2\right)\dot{X}^2\right]\!XM}{2R^5}\!-\!\frac{3Y^3\dot{X}\dot{Y}M}{R^5}
\!-\!\frac{\left[10R^2\dot{t}^2\!-\!(2X^2\!+\!9Y^2)\dot{X}^2\right]\!XM^2}{2R^6}   \\
&&\nn+\frac{(X^2+4Y^2)X\dot{Y}^2 M}{2R^5}-\frac{\left(5X^2-9Y^2\right)Y \dot{X}\dot{Y}M^2}{2R^6}+\frac{\left[63R^2\dot{t}^2-\left(7 X^2+ 27 Y^2\right)\dot{X}^2\right]XM^3}{4 R^7}  \\
&&\nn-\frac{7X Y^2\dot{Y}^2M^2 }{2R^6}+\frac{\left(13X^2-27Y^2\right)Y\dot{X}\,\dot{Y} M^3}{4R^7}-\frac{\left[\,324\,R^2\,\dot{t}^2-\left(20X^2+81Y^2\right)\dot{X}^2\,\right]X M^4}{8 R^8} ~~~~~~ \\
&&+O(M^4)~,   \label{B-X-4} \\
&&\nn0={\ddot{Y}}+\frac{\left[\,2\,R^2\,\dot{t}^2+\left(4X^2+Y^2\right)\dot{X}^2\,\right]Y M}{2R^5}-\frac{3\,X^3\,\dot{X}\,\dot{Y}M}{R^5}-\frac{\left(10\,R^2\,\dot{t}^2+7X^2\dot{X}^2\right)Y\!M^2}{2R^6}  \\
&&\nn-\,\frac{\left(\,5\,X^2+2\,Y^2\,\right)Y\,\dot{Y}^2M}{2R^5}+\frac{\left(\,9\,X^2-5\,Y^2\,\right)X\dot{X}\,\dot{Y}M^{\,2}}{2R^6}+\frac{\left(\,63\,R^2\,\dot{t}^2+20\,X^2\,\dot{X}^2\,\right)YM^3}{4R^7}  \\
&&\nn+\,\frac{\left(9\,X^2+2\,Y^2\right)Y\,\dot{Y}^2\,M^2}{2R^6}-\frac{\left(27X^2-13Y^2\right)X\dot{X}\,\dot{Y}M^3}{4R^7}-\frac{\left(324\,R^2\,\dot{t}^2+61 X^2\dot{X}^2\right)\!YM^4}{8R^8} ~~~~~~ \\
&&+O(M^4)~.  \label{B-Y-4}
\end{eqnarray}}
It can be seen that Eqs.~\eqref{B-t},~\eqref{B-X-4} and~\eqref{B-Y-4} are the same as Eqs.~\eqref{MEt} - \eqref{MEy} given above.

\end{document}